\documentclass[a4paper,nobibnotes,nofootinbib,notitlepage,superscriptaddress]{revtex4-1}

\usepackage{amssymb}
\usepackage{amsmath}
\usepackage{epsfig}
\usepackage{graphicx}
\usepackage{subfloat}
\usepackage{floatrow}
\usepackage{subfigure}
\usepackage{array}
\usepackage{lineno}
\usepackage{adjustbox}
\usepackage{epstopdf}
\usepackage[utf8]{inputenc}
  \usepackage{csquotes}
\DeclareGraphicsExtensions{.eps}

\newcommand{\be}{\begin{equation}}
\newcommand{\ee}{\end{equation}}
\newcommand{\bea}{\begin{eqnarray}}
\newcommand{\eea}{\end{eqnarray}}

\newcommand{\pom}{\mathbb{P}}
\newcommand{\reg}{\mathbb{R}}

\newcommand{\comm}[1]{}

\begin{document}

\title{Diffractive di-jet production at the LHC with a Reggeon contribution}

\author{C. Marquet}\email{cyrille.marquet@polytechnique.edu}
\affiliation{Centre de Physique Th\'eorique, \'Ecole Polytechnique, CNRS, Universit\'e Paris-Saclay, F-91128 Palaiseau, France}

\author{D.E. Martins}\email{dan.ernani@gmail.com}
\affiliation{Centre de Physique Th\'eorique, \'Ecole Polytechnique, CNRS, Universit\'e Paris-Saclay, F-91128 Palaiseau, France}
\affiliation{Instituto de F\'isica - Universidade do Estado do Rio de Janeiro, Rio de Janeiro 20550-900, RJ, Brazil}

\author{A.V. Pereira}\email{antonio.vilela.pereira@cern.ch}
\affiliation{Instituto de F\'isica - Universidade do Estado do Rio de Janeiro, Rio de Janeiro 20550-900, RJ, Brazil}

\author{M. Rangel}\email{murilo.rangel@cern.ch}
\affiliation{Instituto de F\'isica - Universidade Federal do Rio de Janeiro, Rio de Janeiro 21941-901, RJ, Brazil}

\author{C. Royon}\email{christophe.royon@cern.ch}
\affiliation{University of Kansas, Lawrence, KS 66045, USA}

\begin{abstract}
We study hard diffractive scattering in hadron-hadron collisions including, on top of the standard Pomeron-initiated processes, contributions due to the exchange of Reggeons. Using a simple model to describe the parton content of the Reggeon, we compute di-jet production in single diffractive and central diffractive events. We show that Reggeon contributions can be sizable at the LHC, and even sometimes dominant, and we identify kinematic windows in which they could be experimentally studied. We argue that suitable measurements must be performed in order to properly constrain the model, and be able to correctly account for Reggeon exchanges in the analysis of the many hard diffractive observables to be measured at the LHC.
\end{abstract}

\maketitle

\section{Introduction}

Hard diffractive events in hadron-hadron collisions were first observed at the Tevatron \cite{Abachi:1994hb,Abe:1994de} more than 20 years ago, nevertheless the QCD dynamics at play has yet to be fully understood. In spite of the large transfer of transverse momentum involved in such processes, a satisfactory weak-coupling description remains elusive, and one has to settle for phenomenological models.
To estimate cross-sections of hard processes in single diffractive dissociation (when one hadron escapes the collision intact) and central diffractive dissociation (when both hadrons escape the collision intact), a modern version of the resolved-Pomeron model \cite{Ingelman:1984ns} is being widely used.

This model describes hard diffractive scattering in the following way: hadrons scatter through the exchange of a colorless objet called the Pomeron, which carries a longitudinal momentum fraction denoted $\xi$ and a four-momentum squared denoted $t$. Then, imitating what happens in collinear factorization, a long-distance/short-distance separation of the Pomeron-induced subprocesses is assumed, into perturbative partonic cross-sections and non-perturbative parton distribution functions (pdfs) of the Pomeron, that depend on the parton fractional longitudinal momentum $\beta$, and on the hard scale of the problem $\mu^2$.

The motivation for this model comes from the fact that, in electron-hadron collisions, the diffractive part of the deep inelastic scattering (DIS) cross-section does obey collinear factorization \cite{Collins:1997sr}. The further factorization of the diffractive parton densities $f_{a/h}^D$ into a Pomeron flux $\Phi_{\pom/h}(\xi,t)$ and Pomeron parton distributions $f_{a/\pom}(\beta,\mu^2)$ is an assumption, called Regge factorization, which is accurate and routinely used in QCD fits of diffractive DIS data, from which $f_{q/\pom}$, $f_{g/\pom}$ and $\Phi_{\pom/h}$ are extracted.

When imported to hadron-hadron collisions, such factorization does not apply for diffractive processes, even at very large momentum scales, as shown by comparisons to Tevatron data \cite{Affolder:2000vb}. The presence of additional soft interactions between the colliding hadrons, which may fill the rapidity gap(s), is the standard interpretation of this factorization breaking, and there are empirical indications that it can be compensated by an overall factor, called the gap survival probability, roughly independent of the details of the hard process. In fact, the phenomenology of this factor is still a topic of intense debate~\cite{sgap1,sgap2,sgap3,sgap4} and it represents the last ingredient of the resolved-Pomeron model.

At the LHC, a whole new set of experimental studies has started, in order to provide answers to a number of unsolved questions. Is the gap survival probability only a function of the collision energy, as often assumed? Does one need a different factor for single diffraction and central diffraction? Is the quark and gluon composition of the Pomeron extracted from HERA data compatible with LHC measurements? In this letter, we would like to study a different aspect that has not been investigated extensively: the possibility that the diffractive scattering happens through the exchange of a Reggeon, as opposed to a Pomeron. As a matter of fact, quality fits to diffractive DIS data do require that both Pomerons and Reggeons contribute to the diffractive pdfs: $f_{a/h}^D\!=\!\Phi_{\pom/h}f_{a/\pom}\!+\!\Phi_{\reg/h}f_{a/\reg}$. 
The differences between the two contributions resides in the $\xi$ and $t$ dependence of their fluxes. Reggeon exchange matters mostly at high $\xi$, notably for $\xi>0.1$ and the shape of the $t$ distribution is also different, showing a less steep decrease than in the Pomeron case. 

At the LHC, when large diffractive masses are considered - which is the case in a number of studies (see for instance \cite{N.Cartiglia:2015gve}) - such large values are easily reached and one may therefore wonder about the importance of the Reggeon contribution. 
Previous studies have shown that the Reggeon contribution is not always negligible~\cite{Luszczak:2014mta,Luszczak:2014cxa,Luszczak:2016csq} but this is not always taken into account by standard codes. Our goal in this work is to illustrate, within a very simple model where the parton content of the Reggeon is obtained from the pion structure function, that indeed the Reggeon contribution cannot be safely neglected, and that for processes where both protons escape the collision intact, a double-Reggeon exchange can even dominate over a double-Pomeron exchange.

The plan of the letter is as follows. In Section 2, we present more details about the resolved-Pomeron model for hard diffraction in hadron-hadron collisions, we explain its implementation into the Forward Physics Monte Carlo (FPMC) program \cite{fpmc} that we shall utilize, and we outline our subsequent analysis of diffractive di-jet production, the process we have chosen to consider. In Section 3, we present our results, when the Reggeon contribution is included, for both single and central diffractive di-jets. Section 4 is devoted to conclusions and outlook.

\section{Hard diffractive processes with Reggeon exchanges}

\subsection{Resolved Pomeron model supplemented with Reggeons}

\begin{figure}
\includegraphics[scale=0.37]{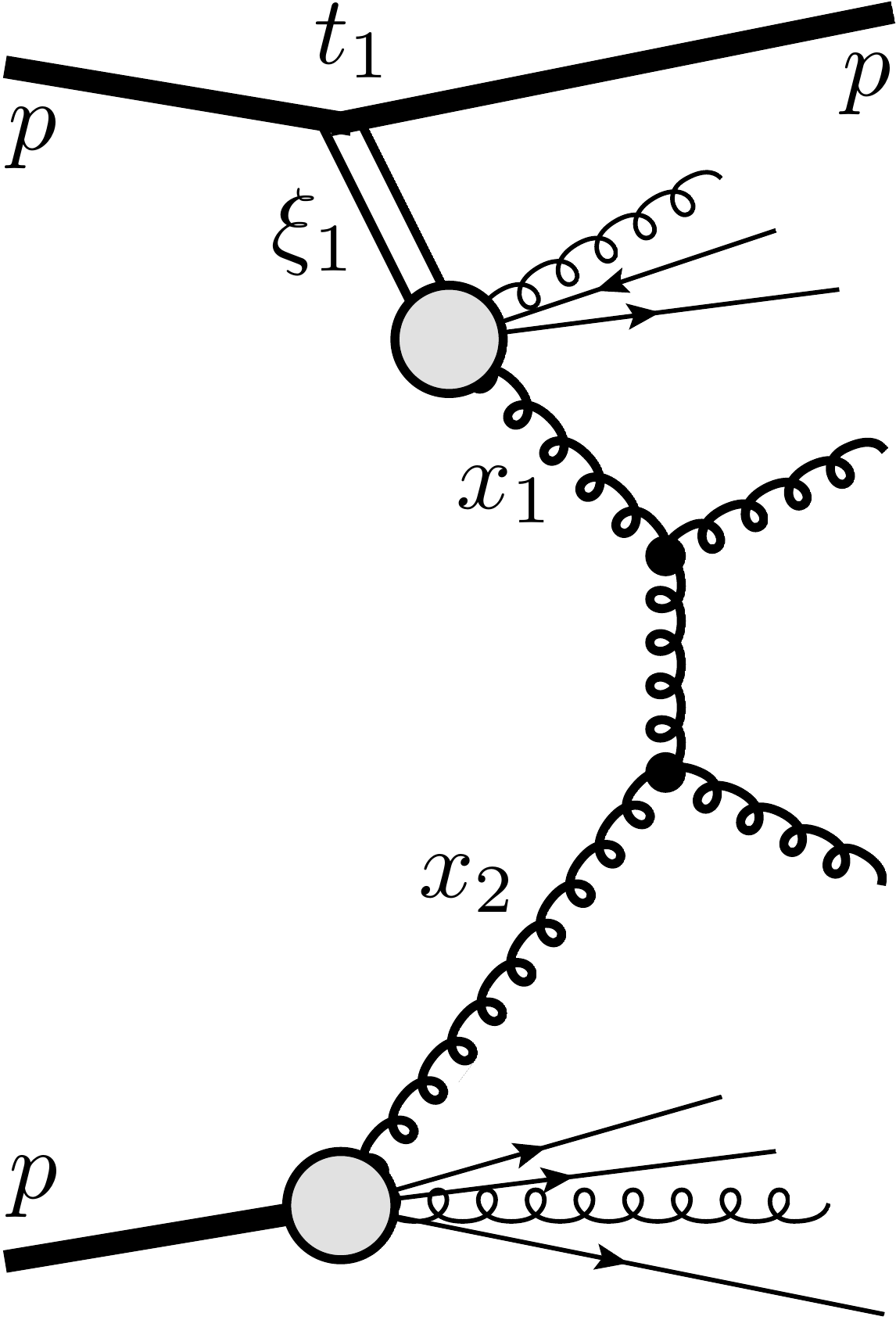}
\hspace{2.5cm}
\includegraphics[scale=0.37]{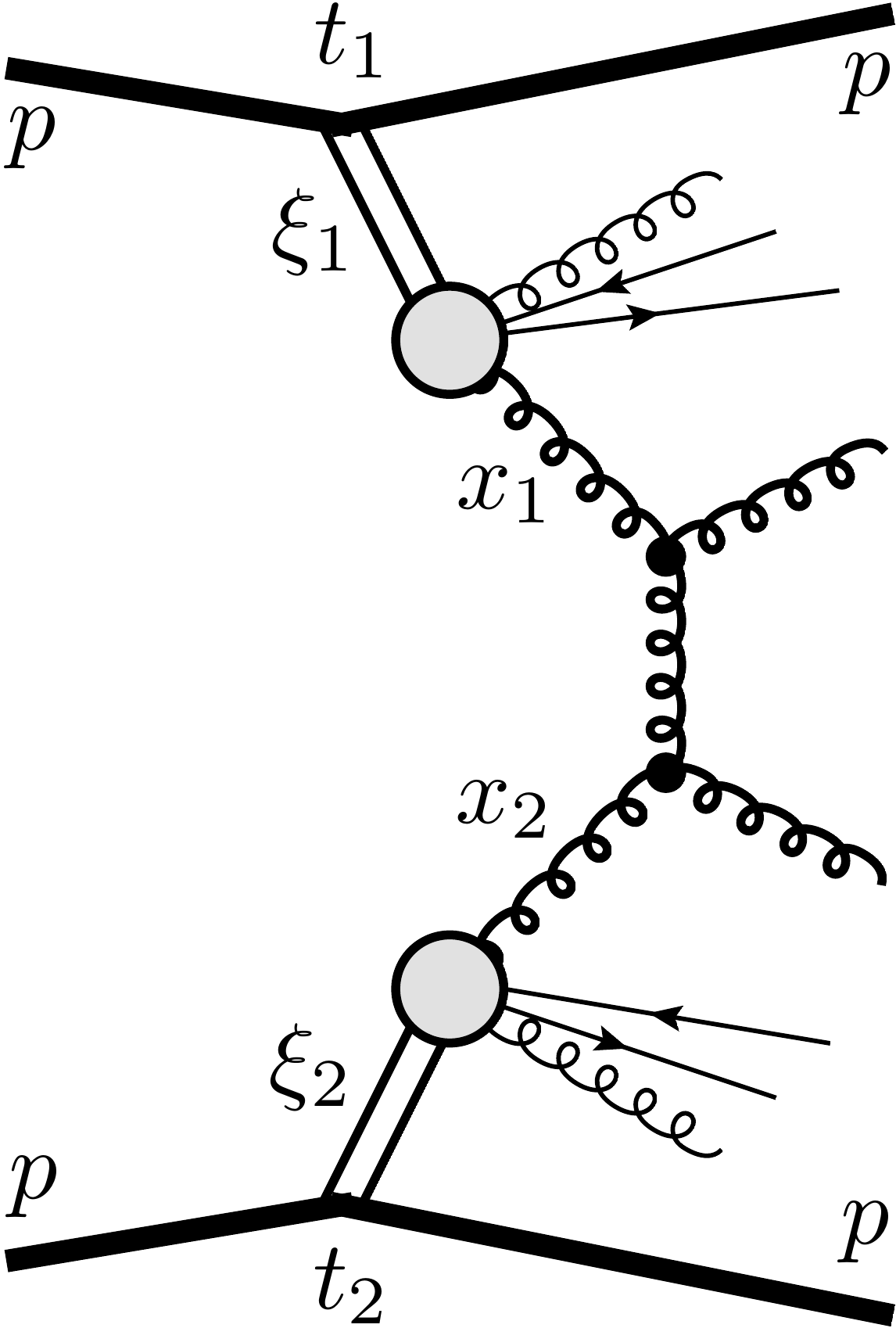}
\caption{Leading-order diagrams for di-jet production in single-diffractive events (left) and central-diffractive events (right) in proton-proton collisions. Intact protons can scatter through the exchange of either a Pomeron ($\pom$) or a Reggeon ($\reg$).}
\label{fig0}
\end{figure}

The resolved-Pomeron model is a long-distance/short-distance collinear factorization framework commonly used to calculate hard single-diffractive (SD) and central-diffractive (CD) processes. In this work we focus on di-jet production at the LHC. Typical leading-order (LO) diagrams for this process are pictured in Fig.~\ref{fig0}, and the cross-section in the resolved-Pomeron model reads:
\bea
d\sigma^{pp\to pJJX}={\cal S}_{SD}\ &&\sum_{a,b} \int f^D_{a/p}(\xi_1,t_1,\beta_1,\mu^2)f_{b/p}(x_2,\mu^2)\
\otimes d\hat\sigma^{ab\to JJX}\label{colfact1}
\eea
\bea
d\sigma^{pp\to pJJXp}={\cal S}_{CD}\ &&\sum_{a,b} \int f^D_{a/p}(\xi_1,t_1,\beta_1,\mu^2)f^D_{b/p}(\xi_2,t_2,\beta_2,\mu^2)\
\otimes d\hat\sigma^{ab\to JJX}\label{colfact2}
\eea
where $d\hat\sigma$ is the short-distance partonic cross-section, which can be computed
order by order in perturbation theory (provided the transverse momentum of the jets is
sufficiently large). $f_{a/p}$ denotes the standard proton parton distributions functions (pdfs) while $f^D_{a/p}$ denotes the diffractive ones.
These are non-perturbative objects, however their evolution with the factorization scale $\mu$ is perturbative and given by the
Dokshitzer-Gribov-Lipatov-Altarelli-Parisi~\cite{dglap} evolution equations (in the following $\mu$ is set to the transverse momentum of the leading jet). In Eq.~\eqref{colfact1} and~\eqref{colfact2} and in Fig.~\ref{fig0}, the variables $\xi_i$ and $t_i$ for the intact protons denote their fractional energy loss and the four-momentum squared transferred in the collision, respectively. The convolution is done over the longitudinal momentum fractions of the partons $a$ and $b$ with respect to the incoming protons, namely $x_1$ and $x_2$, respectively. In the case of intact protons, it is common to use instead $\beta_i \equiv x_i/\xi_i$, the longitudinal momentum fraction of the parton with respect to the exchanged Pomeron or Reggeon.

Formulae \eqref{colfact1} and \eqref{colfact2} are reminiscent of the collinear factorization obeyed for inclusive processes. However, it is known that hard diffractive cross-sections in hadronic collisions do not factorize in such a way, due to possible secondary soft interactions between the colliding hadrons which can fill the rapidity gaps. In the resolved-Pomeron model, the so-called gap survival probabilities ${\cal S}_{SD}$ and ${\cal S}_{CD}$ act as corrections to collinear factorization in order to account for the effects of the soft interactions. Since those happen on much longer time scales compared to the hard process, they are modeled by an overall factor, function of the collision energy only~\cite{sgap1,sgap2}.

\subsection{Pomeron and Reggeon parton-content}

In our computations, we shall use the diffractive pdfs $f^D_{a/p}$ extracted from HERA data \cite{Aktas:2006hy} for diffractive DIS, a process for which collinear factorization does hold. They are obtained by means of a next-to-leading-order (NLO) QCD fit, in which they are decomposed into Pomeron and Reggeon fluxes $\Phi_{\pom,\reg/p}$ and their corresponding parton distribution functions $f_{a/\pom,\reg}$ which depict the partonic structure of the exchanged color singlet objects:
\be
f^D_{a/p}(\xi,t,\beta,\mu^2) =  \Phi_{\pom/p}(\xi,t) f_{a/\pom}(\beta,\mu^2) + n_{\reg} \Phi_{\reg/p}(\xi,t) f_{a/\reg}(\beta,\mu^2) \quad \mbox{with}
\quad \Phi_{\pom,\reg/p}(\xi,t) =  A_{\pom,\reg}\ \frac{e^{B_{\pom,\reg}t}}{\xi^{2\alpha_{\pom,\reg}(t)-1}}\ .
\label{dpdfs}
\ee
The diffractive slopes $B_{\pom,\reg}$, and the Regge trajectories $\alpha_{\pom,\reg}(t)=\alpha_{\pom,\reg}(0)+t\alpha'_{\pom,\reg}$ are given in Table \ref{fitvalues} for two different fits (known as $A$ and $B$). 
The flux normalizations $A_{\pom,\reg}$ are chosen such that $\xi\times\int_{t_{\rm{min}}}^{t_{\rm{max}}}dt\ \Phi_{\pom,\reg/p}(\xi,t)=1$ at $\xi=0.003$, with $t_{\rm{min}}=-1$ GeV$^2$ and $t_{\rm{max}}=-m_p^2\xi^2/(1\!-\!\xi)$ ($m_p$ denotes the proton mass). 
The factor $n_{\reg}$ is an extra normalization to the Reggeon contribution.
 
The Pomeron pdf $f_{a/\pom}$ is obtained from fits to H1 data~\cite{Aktas:2006hy}. 
The Pomeron structure is well constrained by those fits, which clearly show that its parton content is gluon dominated. 
By contrast, the HERA data do not constrain $f_{a/\reg}$. The Reggeon contribution is however needed in order to obtain a quantitative description of the high-$\xi$ measurements. 
Following the description in Ref.~\cite{Aktas:2006hy} we treat the Reggeon contribution as an exchange of a quark-antiquark pair and take $f_{a/\reg}$ as the pion structure function. 

\begin{table}[b]
\begin{center}
\begin{tabular}{|c|c|c|c|c|c|}
\hline
\ \textbf{Fits}\, &$\ \Phi_{\pom,\reg/p}\, $&$\alpha(0)$ & $\alpha'$ & $n_{\reg}$ & $B$   \\ \hline &&&&&\\
&$\pom$ & $1.118\pm 0.008$ & $0.06_{-0.06}^{+0.19}$ GeV$^{-2}$ & - & $5.5_{-2.0}^{+0.7}$ GeV$^{-2}$  \\ A &&&&& \\
&$\reg$ & $0.50\pm 0.10$ & $0.3^{+0.6}_{-0.3}$ GeV$^{-2}$ & $(1.7\pm 0.4) \times 10^{-3}$  & $1.6_{-1.6}^{+0.4}$ GeV$^{-2}$ \\ &&&&& \\ \hline &&&&& \\
&$\pom$ & $1.111\pm 0.007$ & $0.06_{-0.06}^{+0.19}$ GeV$^{-2}$ & - & $5.5_{-2.0}^{+0.7}$ GeV$^{-2}$ \\ B &&&&& \\
&$\reg$ & $0.50\pm 0.10$ & $0.3^{+0.6}_{-0.3}$ GeV$^{-2}$ & $(1.4\pm 0.4) \times 10^{-3}$  & $1.6_{-1.6}^{+0.4}$ GeV$^{-2}$ \\ &&&&& \\ \hline
\end{tabular}
\end{center}
\caption{
Parameters of the Pomeron and Reggeon fluxes as described in the text, from Ref.~\cite{Aktas:2006hy}. 
}
\label{fitvalues}
\end{table}

The Reggeon contribution is often not implemented in hard diffraction studies in hadron-hadron collisions, and measurements at the LHC will allow to test the validity of this assumption. 
The Reggeon contribution to the diffractive pdfs $f^D_{a/p}$ is important only at large values of $\xi$~\cite{Aktas:2006hy,Aktas:2006hx}.
It is routinely disregarded, and subsequently the theoretical description from \eqref{colfact1} and \eqref{colfact2} was dubbed the resolved-Pomeron model. But of course it can be supplemented with resolved Reggeons, and it should be as we will argue below. 

Similarly, this is the reason why what we call central diffractive events in this work is usually referred to as double-Pomeron-exchange events in the literature. Since double-Reggeon exchanges or mixtures of Pomeron and Reggeon exchanges are also possible, we choose to use the terminology \enquote{central diffractive}. 
The central diffractive di-jet final states considered in this work are not exclusive since they contain the so-called Pomeron or Reggeon remnants $X$, that are made of soft particles accompanying the production of the hard di-jet system. They reduce the rapidity gaps when compared to the exclusive case, but they do not fill them entirely.

\subsection{Hard diffractive di-jet analysis with FPMC}

The above theoretical description of hard diffractive processes in hadron-hadron collisions, in which one or both hadrons remain intact, is implemented by the FPMC generator that we shall employ to perform our analysis. The parton-level matrix elements are imported from HERWIG \cite{herwig} routines and calculated at LO, while the NLO fit B is adopted for the diffractive pdfs~\footnote{We decided to use LO partonic cross-sections convoluted with NLO pdfs due to the lack of an updated LO diffractive pdf.}.  For proton tagging at the LHC, a region of $\xi<0.17$ for both protons is chosen for a center of mass energy of 13 TeV, as well as a lower cut of $0.15$ GeV for their transverse momenta \cite{Trzebinski:2014vha}.
In general, the lower boundary for the $\xi$ values depends on the minimum mass of the diffractive system and is thus related to the jet transverse momenta. For the acceptance we have chosen the lower $\xi$ boundary is approximately $10^{-5}$.

For the di-jet system, we apply at parton level a transverse momentum cut on $p_T\!>\!5$ GeV, and a pseudo-rapidity cut of $|\eta|\!<\!5$. The jets are reconstructed using the FastJet~\cite{fastjet:2012} package and the anti-kt algorithm, with a value of 0.4 for the jet radius, and a 10 GeV threshold for the transverse momenta. Then, the selection criteria requires at least two jets with $p_T$ larger than 20 GeV, and the two highest transverse momentum jets tagged with $p_{T}(j_{1})\!>\!p_{T}(j_{2})$. The di-jet mass fraction is defined as $R_{JJ}\!=\!m_{JJ}/M$, i.e. the ratio of the invariant mass of the di-jet system to the invariant mass of the whole diffractive
final state, $M\!=\!\sqrt{\xi s}$ and $M\!=\!\sqrt{\xi_{1}\xi_{2}s}$ for single and central diffraction, respectively.

Experimentally, the di-jet mass fraction is a good variable for identifying, and for our purpose excluding, possible exclusive di-jet events. In such events, the di-jet mass is essentially equal to the mass of the central system 
because no Pomeron (or) Reggeon remnants are present, and if the jet definition is such that little is left outside the cones, then the presence of an exclusive event would manifest itself as an excess towards $R_{JJ} \approx 1$.
This observation of exclusive events does not depend on the overall normalization of the event distribution, which might be strongly dependent on the detector simulation and acceptance of the roman pot detectors \cite{Kepka:2007nr}.

Finally, the histograms are normalized according to the relation $(\sigma \times \mathcal{L})/N_{gen}$, and our predictions are presented for an integrated luminosity of $1$~pb$^{-1}$ which represents the expected data to be collected in high-$\beta^*$ low pile-up runs at the LHC. Note that for the gap survival probabilities, we have assumed a constant value for Pomerons and Reggeons $S_{SD}=S_{CD} \!\simeq\! 0.03$. There have been several attempts to estimate those probabilities~\cite{Khoze:2000cy,Khoze:2008cx,Kaidalov:2001iz,Bartels:2006ea,Luna:2006qp,Frankfurt:2006jp,Gotsman:2007ac,Gotsman:2011xc,Achilli:2007pn,sgap1,sgap2,sgap3,sgap4}, but the actual values are rather uncertain.
The chosen value of the gap survival factor can be considered a lower limit given the recent available experimental results~\cite{Chatrchyan:2012vc,Aad:2015xis}.

\begin{figure}[t]
\includegraphics[width=0.47\textwidth]{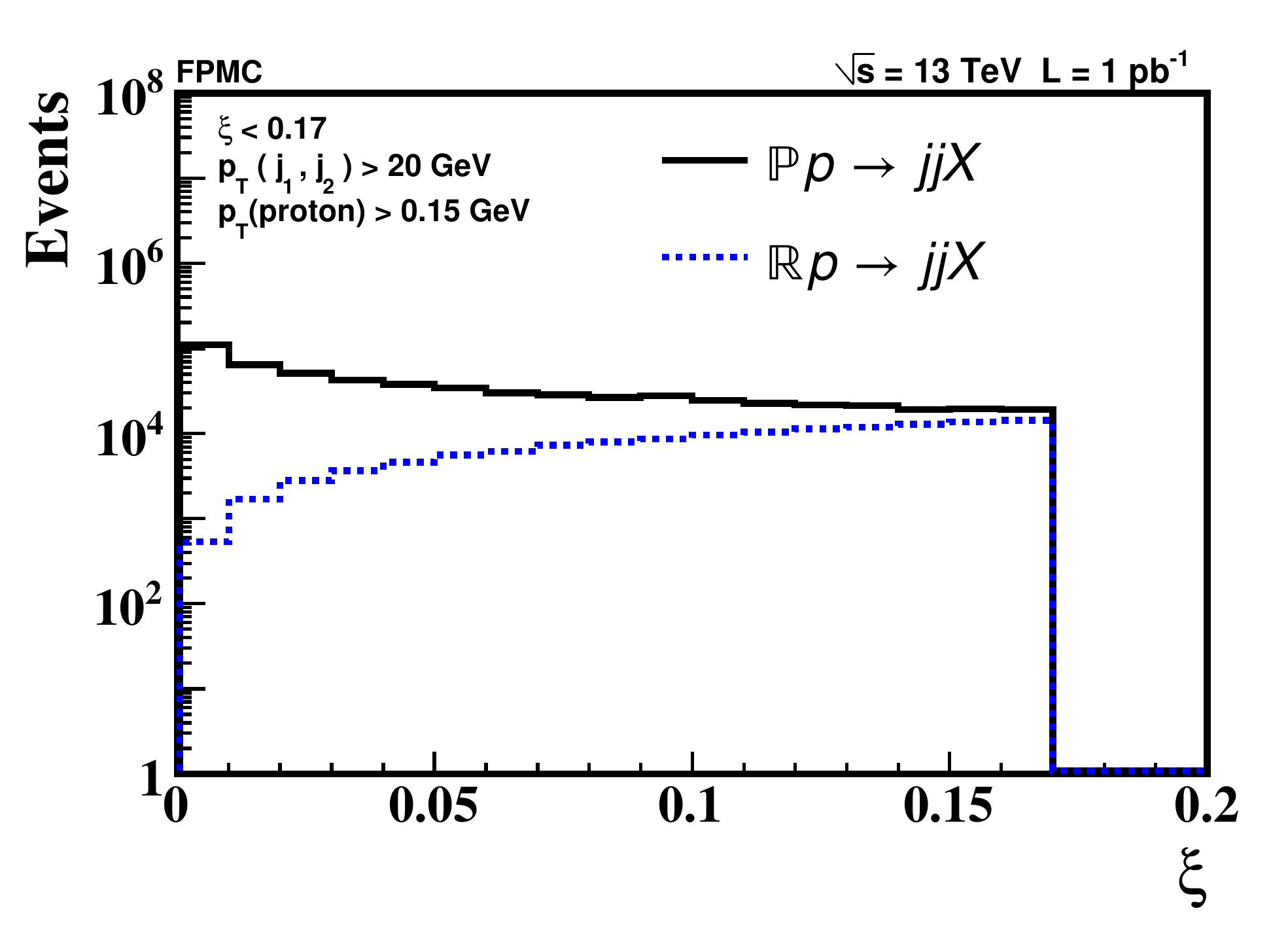}
\includegraphics[width=0.47\textwidth]{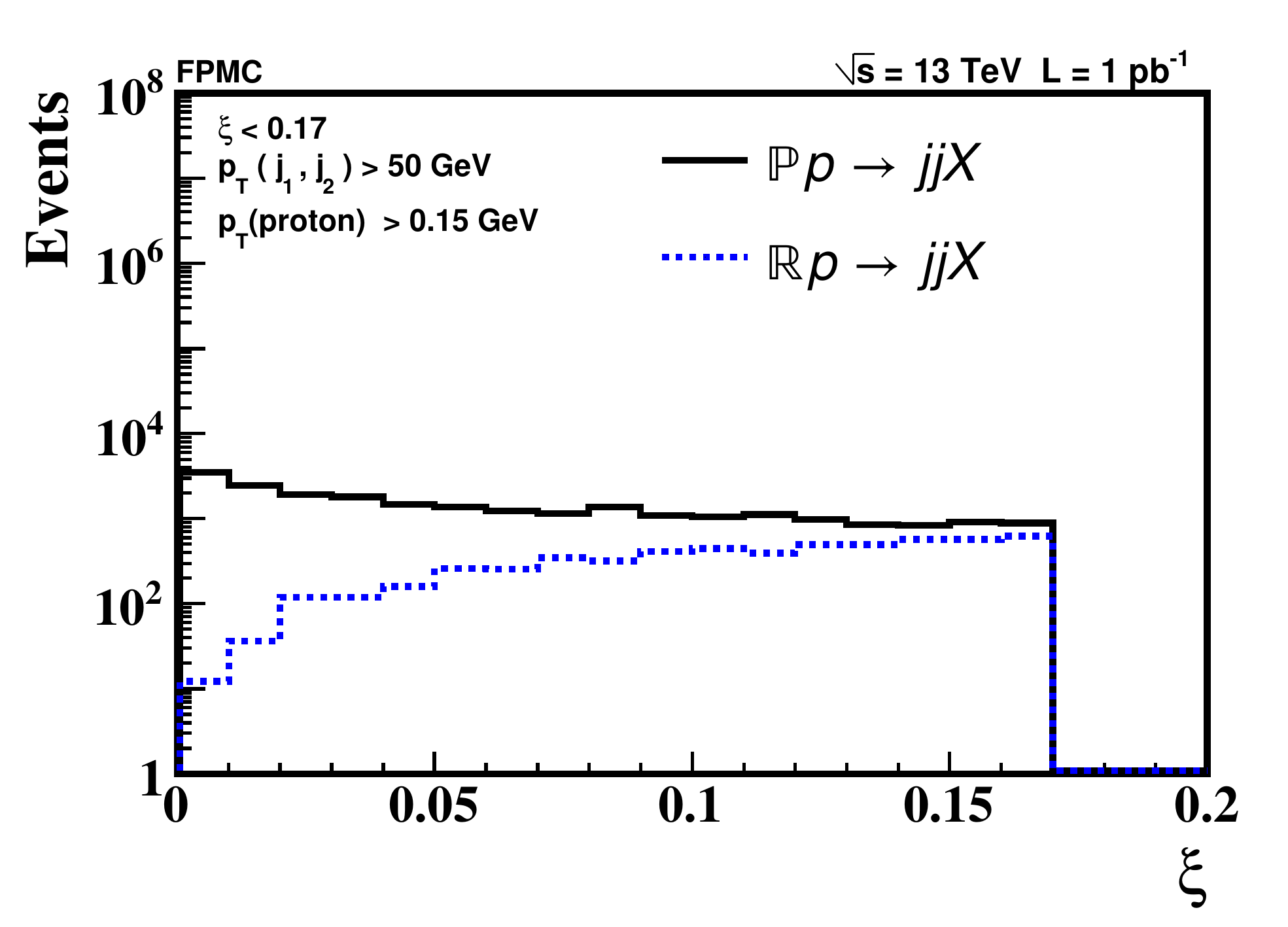}
\caption{Number of single diffractive di-jet events as a function of $\xi$ for p$_{T}$(proton)$>$ 0.15 GeV assuming either a Pomeron exchange (solid lines) or a Reggeon exchange (dashed lines), for $p_{T}(j_{1,2})>20$ GeV (left plot) or $p_{T}(j_{1,2})>50$ GeV (right plot).}
\label{xi_sd}
\end{figure}

\section{Numerical results for the LHC}

\subsection{Single diffractive di-jets}

The total cross-sections predicted by FPMC at 13 TeV for single diffractive di-jets assuming either a Pomeron exchange ($\pom+p\to jjX$) or a Reggeon exchange ($\reg+p\to jjX $) are $1.51 \times 10^{8}$ pb and $2.3\times 10^{7}$ pb, respectively. These values assume an acceptance of $\xi_1\equiv\xi\leq 0.17$ for the final state intact proton. The $\xi$ distributions are plotted in Figure~\ref{xi_sd} for two different values of minimum jet $p_T$, 20 or 50 GeV. One clearly sees the dominance of the Pomeron exchange at small $\xi$ (the Reggeon contribution can be neglected for $\xi\lesssim0.07$), but also the fact that the Reggeon contribution becomes comparable to it for $\xi\gtrsim0.1$, depending slightly on the jet $p_T$ cut.

\begin{figure}[t]
\includegraphics[width=0.47\textwidth]{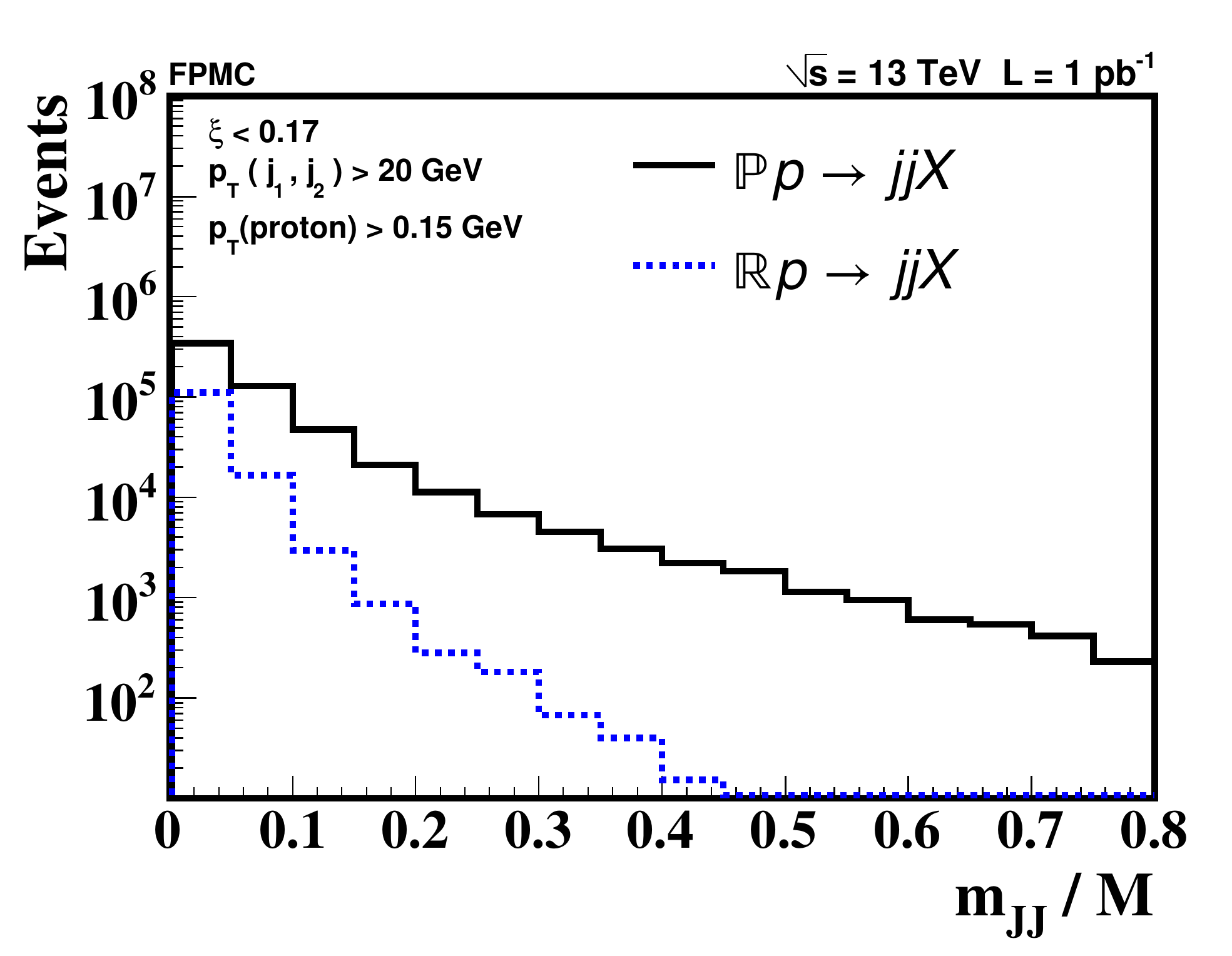}
\includegraphics[width=0.47\textwidth]{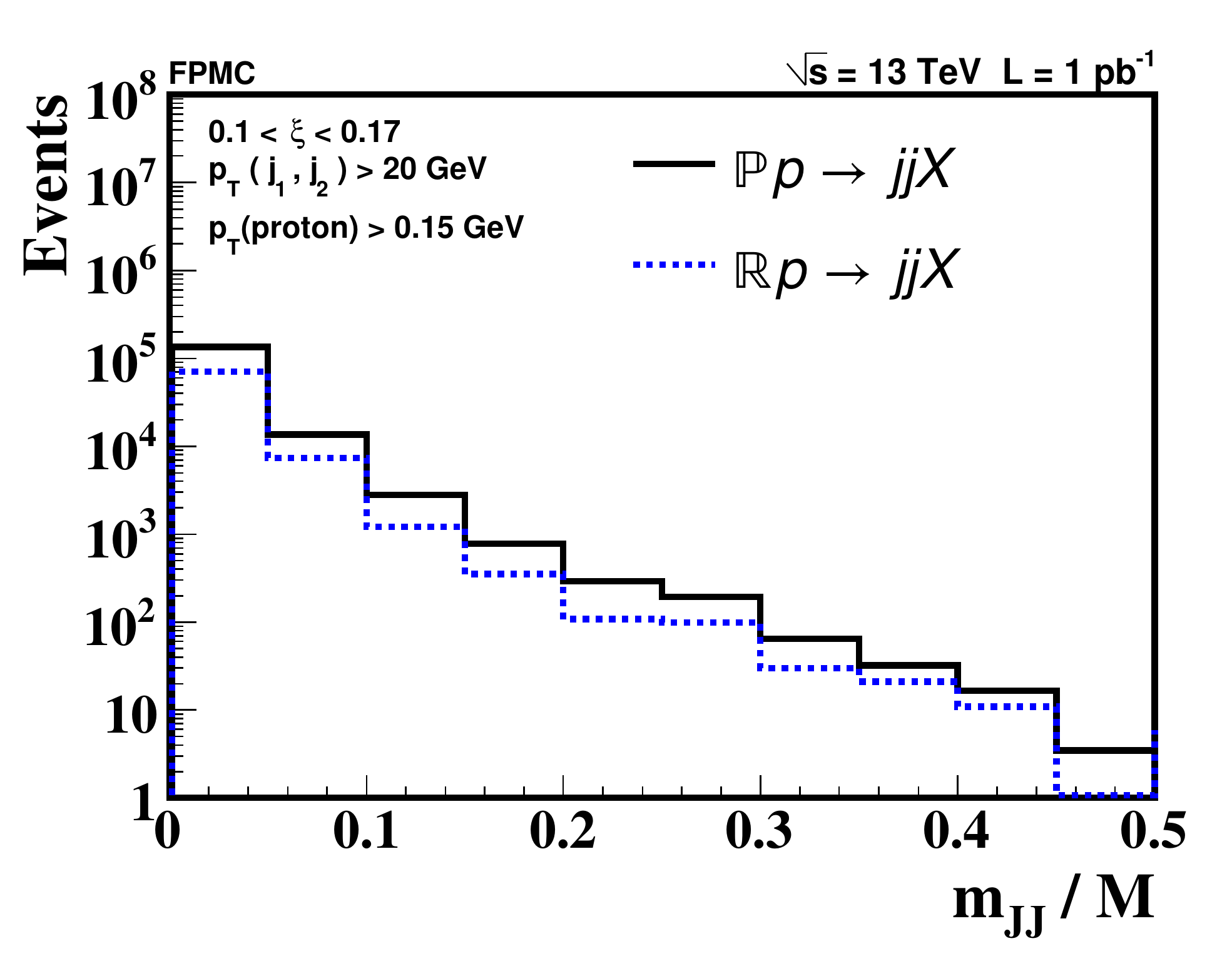}
\caption{Di-jet mass fraction distribution in single diffraction assuming either a Pomeron exchange (solid lines) or a Reggeon exchange (dashed lines), for
$p_{T}(j_{1,2})>20$ GeV, p$_{T}$(proton) $>$ 0.15 GeV and for $\xi<0.17$ (left) or $0.1<\xi<0.17$ (right).}
\label{dmfsdpt04}
\end{figure}

This is confirmed by Table~\ref{neventssd_ptcuts} where the number of events are displayed for three different $\xi$ ranges: no minimum $\xi$ cut, $\xi>0.015$ and $\xi>0.1$. In the latter case, the number of events for the Pomeron and Reggeon contributions have the same order of magnitude. Note that the events for the Pomeron process take into account the uncertainty of the QCD fits at high $\beta$: the gluon density $f_{g/\pom}(\beta,Q^2)$ is multiplied by an uncertainty factor $(1-\beta)^{\nu}$, with $\nu = -0.5$, $0$, or $0.5$ (the default value in FPMC is $\nu=0$). The uncertainty range of the Reggeon contribution is not known. Finally the di-jet mass fraction distributions are displayed in Figure \ref{dmfsdpt04}, for $\xi<0.17$ and $0.1<\xi<0.17$. While the Reggeon exchange can be as important as the Pomeron exchange, there is no kinematical window where it clearly dominates which would allow to experimentally isolate it.

Those findings confirm the expectations, that for single diffractive processes sensitive to $\xi>0.1$, the Reggeon contribution should play a non-negligible role. Therefore, the LHC capabilities should be utilized in order to constrain it better, and improve the theoretical predictions of the various high-mass diffractive studies. We will demonstrate below that central diffractive di-jet production can also be used to constrain the Reggeon contribution.

\begin{table}[h]
\renewcommand{\arraystretch}{1.5}%
\adjustbox{max height=\dimexpr\textheight-1.0cm\relax,max width=\textwidth}{
\begin{tabular}{|p{0.18\linewidth}|p{0.30\linewidth}|p{0.10\linewidth}|p{0.30\linewidth}|p{0.10\linewidth}|}
\hline
\textbf{Process}                                    & $\pom\ p\to jjX$     & $\reg\ p\to jjX$ & $\pom\ p\to jjX$ & $\reg\ p\to jjX$ \\
\hline 
\textbf{Acceptance}                                 &  \multicolumn{2}{c|}{$p_{T}(j_{1},j_{2})>20$ GeV }                                               &   \multicolumn{2}{c|}{$p_{T}(j_{1},j_{2})>50$ GeV } \\
\hline
$\xi_{1,2} <  0.17$  & $6.06 \times 10^{5}$ [$5.85\times 10^{5}$, $6.74\times 10^{5} $] & $1.38 \times 10^{4} $     & $ 2.51  \times 10^{4}$ [$ 2.26 \times 10^{4} $, $2.86 \times 10^{4} $ ]    & $5450 $           \\
$0.015 < \xi_{1,2} < 0.17$           & $4.58 \times 10^{5}$ [$4.53\times 10^{5}$, $5.03\times 10^{5} $] & $1.37  \times 10^{4} $     & $ 1.99 \times 10^{4}$ [$ 1.81\times 10^{4} $, $2.32 \times 10^{4} $ ]    & $5419 $  \\
$0.10 < \xi_{1,2} < 0.17$     & $1.49 \times 10^{5}$ [$1.46\times 10^{5}$, $1.62\times 10^{5} $] & $8.77  \times 10^{4} $     & $6561 $ [$6341$, $8827 $]    &$3521$ \\
\hline
\end{tabular}
}
\caption{Number of single diffractive di-jet events for an integrated luminosity of $1$ pb$^{-1}$, and different kinematical windows. For the Pomeron process, the left values inside the brackets stand for $\nu=0.5$, whereas the right values stand for $\nu=-0.5$.}
\label{neventssd_ptcuts}
\end{table}  

\subsection{Central diffractive di-jets}

In central diffraction, besides the double-Pomeron ($\pom\pom\to jjX$) and double-Reggeon ($\reg\reg\to jjX$) exchanges, there are also cross terms
($\pom\reg,\reg\pom\to jjX$) which makes the approximation of disregarding the Reggeon contributions even more questionable. The total cross-sections predicted by FPMC for proton-proton collisions at 13 TeV for those distinct channels are $ 1.7\times 10^{7}$ pb ($\pom\pom$), $9.1 \times 10^{6}$ pb ($\pom\reg\!+\!\reg\pom$), and $9.03 \times 10^{5}$ pb ($\reg\reg$), again for an acceptance of $\xi_{1,2}\leq 0.17$ for the final state intact protons. These values are in agreement with the prediction that single diffractive cross-sections should be approximately 10 times greater than in the central diffractive case~\cite{Aad:2012pw}.

\begin{figure}[t]
\includegraphics[width=0.47\textwidth]{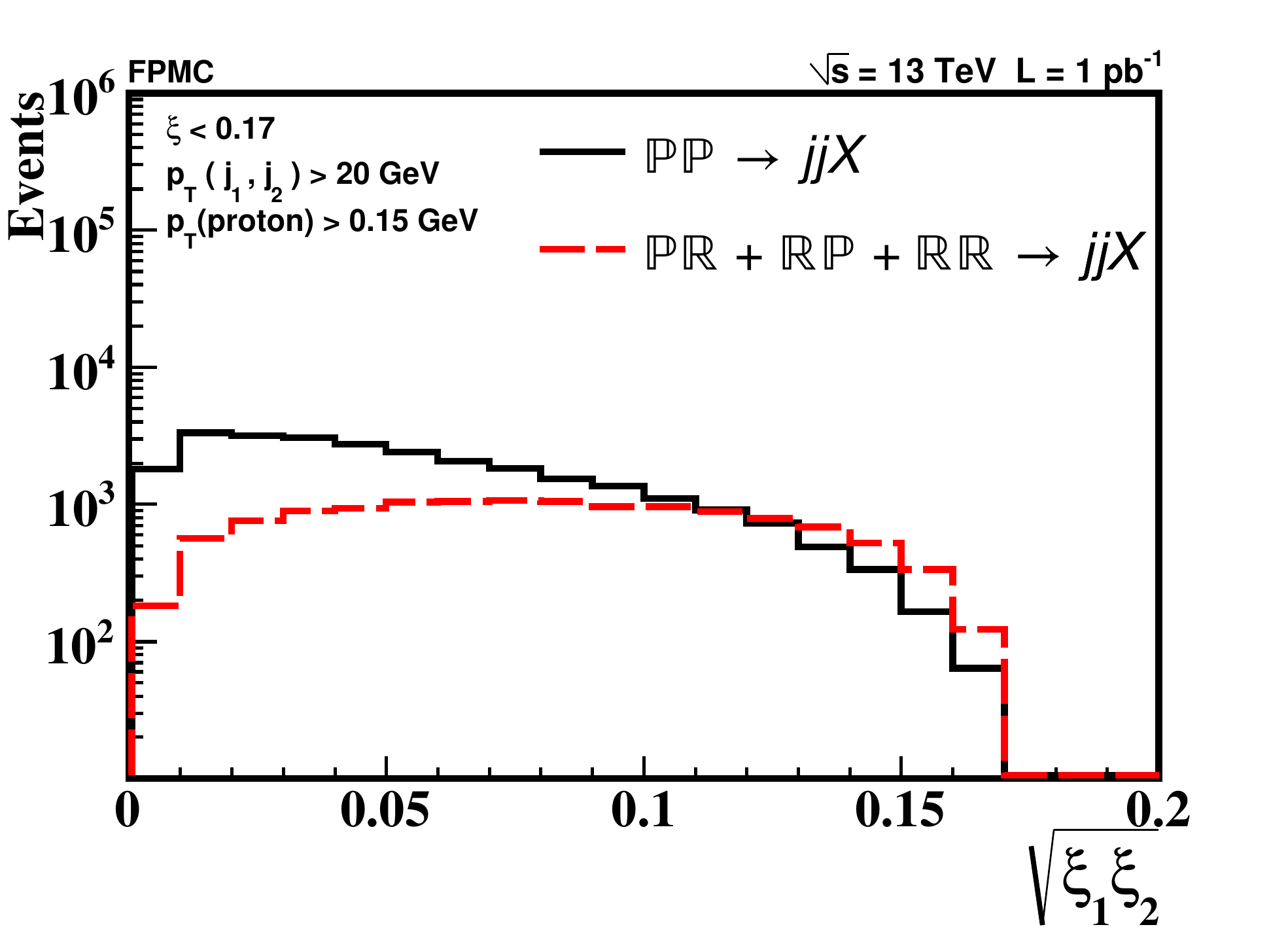}
\includegraphics[width=0.47\textwidth]{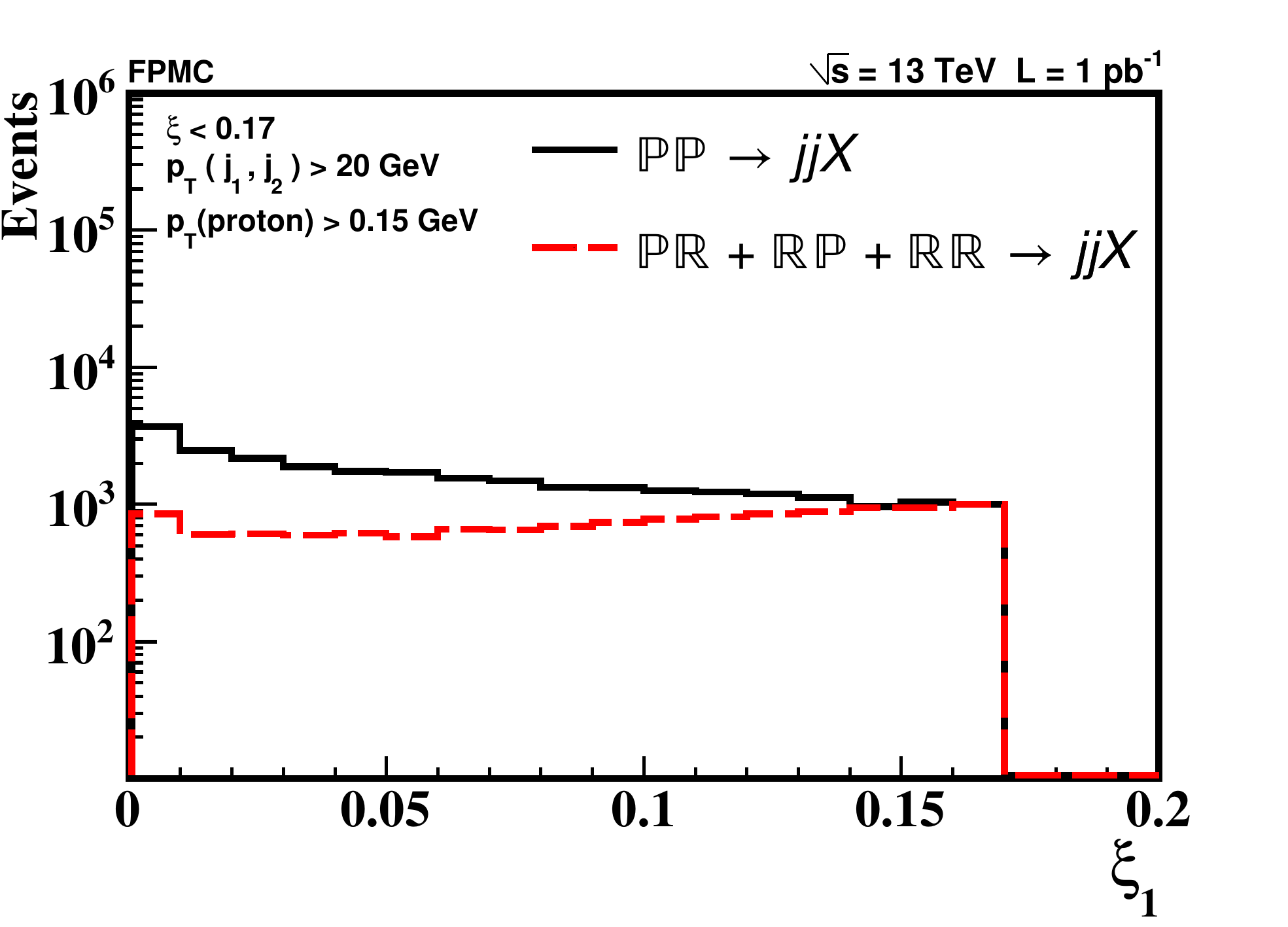}
\caption{Number of central diffractive di-jet events as function of $\sqrt{\xi_1\xi_2}$ (left plot) and $\xi_1$ (right plot) for $p_{T}(j_{1,2}) > 20$ GeV and p$_{T}$(proton) $>$ 0.15 GeV. The solid line stands for the double-Pomeron exchange while the dashed line represents the total Reggeon contribution $(\reg\ \reg$ and $\pom\ \reg+ \reg\ \pom)$.}
\label{centraldiffrative_xi}
\end{figure}

The $\xi$ distributions are plotted in Figure~\ref{centraldiffrative_xi} for a minimum jet $p_T$ of 20 GeV. The Pomeron exchange is still dominant at small values of $\xi_{1,2}$, albeit by a lesser margin than in the single diffractive case, but now the Reggeon contributions dominate for large values of proton momentum loss. They become comparable to the Pomeron one for $\xi_1\sim 0.14$ when $\xi_2$ is integrated in the whole acceptance (Fig.~\ref{centraldiffrative_xi} - right panel).

\begin{figure}[t]
\includegraphics[width=0.49\textwidth]{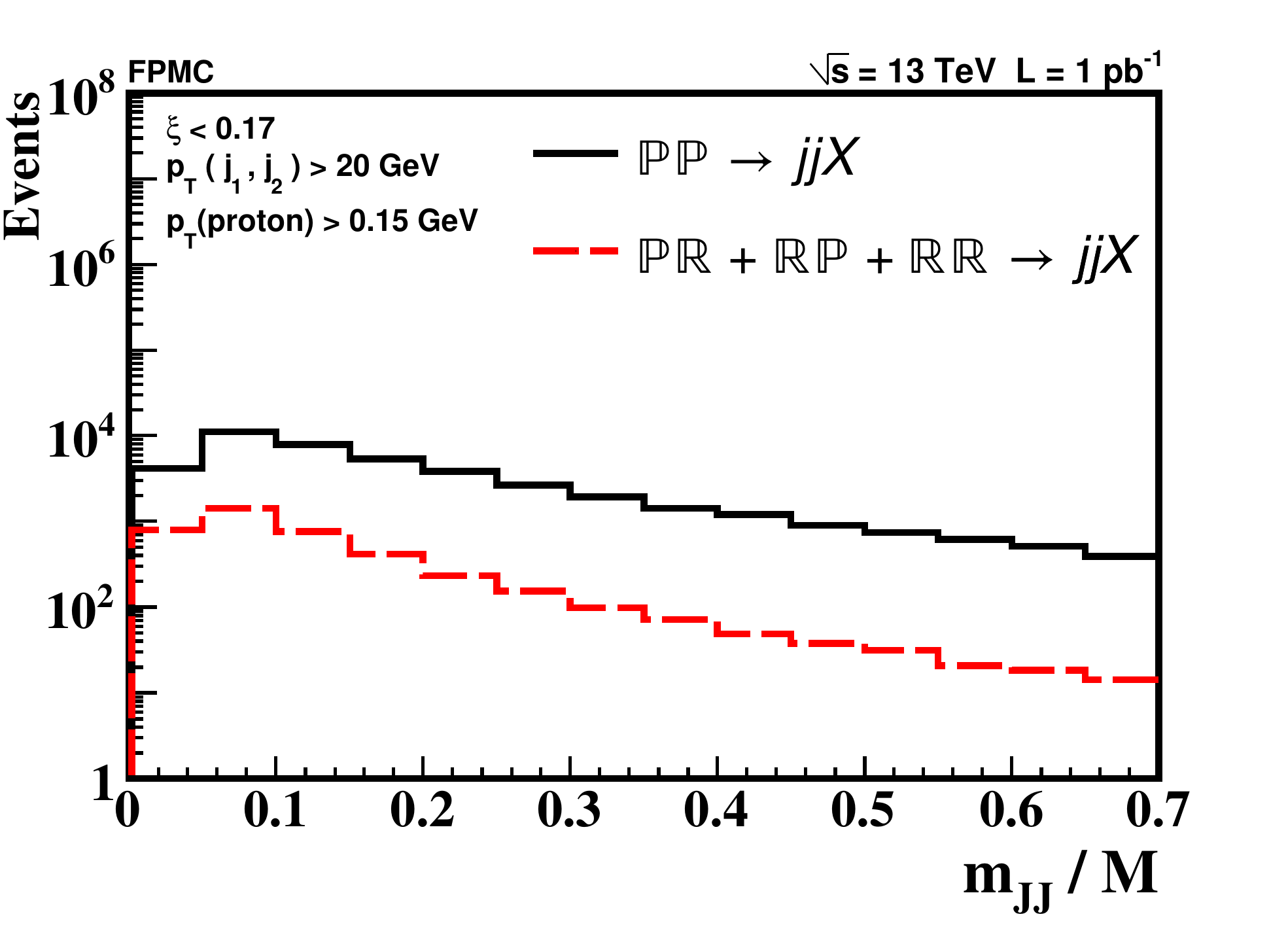}
\includegraphics[width=0.47\textwidth]{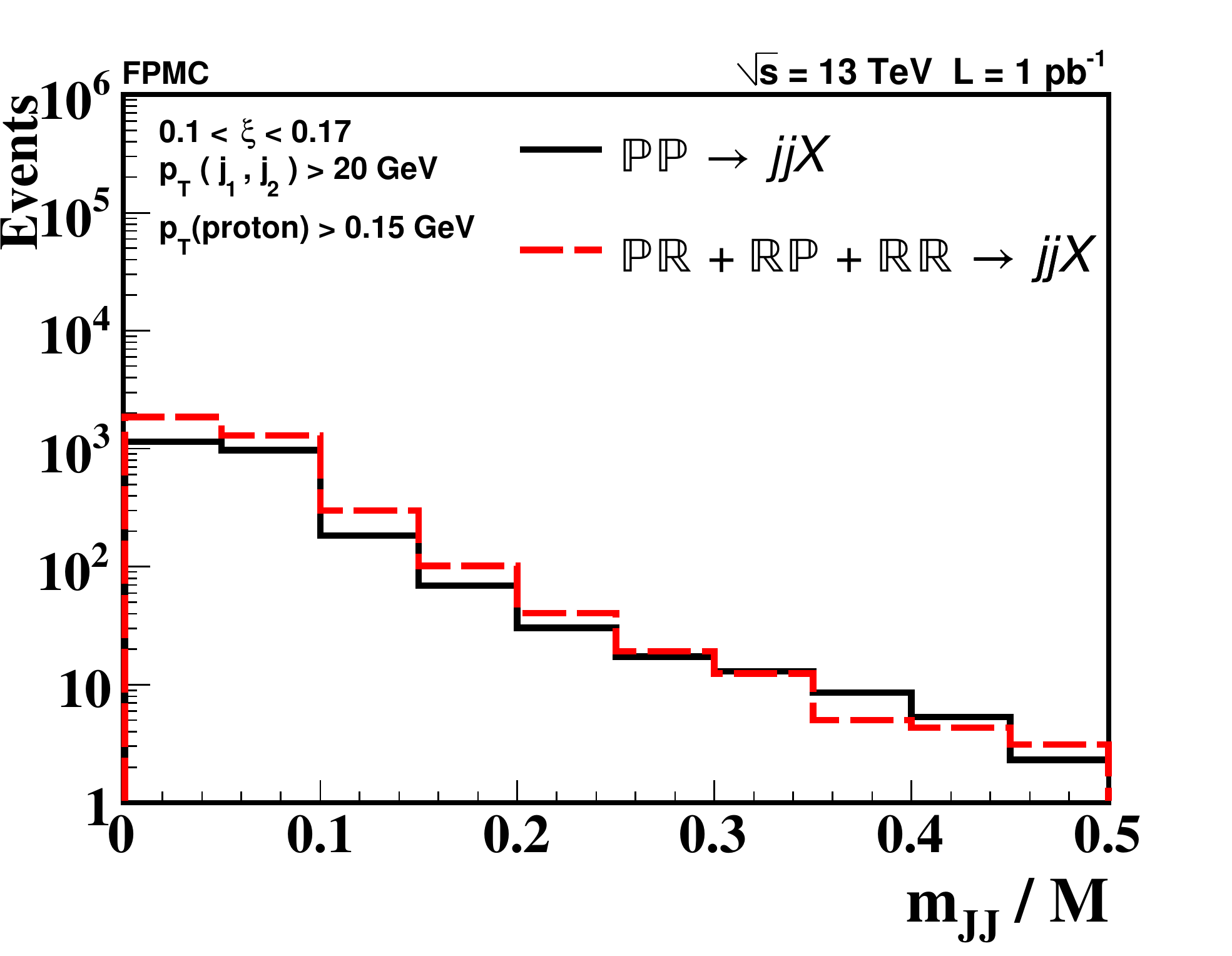}
\caption{Di-jet mass fraction distribution in central diffraction for $p_{T}(j_{1,2})>20$ GeV, p$_{T}$(proton) $>$ 0.15 GeV and for $\xi_{1,2}<0.17$ (left plot) or $0.1<\xi_{1,2}<0.17$ (right plot). The solid line stands for the double-Pomeron exchange while the dashed line represents the total Reggeon contribution from the double-Reggeon and the Pomeron-Reggeon exchanges.
}
\label{dmfcentraldiffrative_xi}
\end{figure}

This is confirmed by Table~\ref{neventsptcut1} where the number of events is displayed for the same three different $\xi_{1,2}$ ranges and two different jet $p_T$ cuts considered before. As expected, using a larger minimum jet $p_T$ further enhances the importance of the Reggeon exchange, however in central diffraction the number of events quickly becomes too small and it may not be efficient to use a value greater than 20 GeV. Finally, the di-jet mass fraction distributions are displayed in Figure \ref{dmfcentraldiffrative_xi}, for $\xi_{1,2}\!<\!0.17$ and $0.1\!<\!\xi_{1,2}\!<\!0.17$.

\begin{table}[h]
\renewcommand{\arraystretch}{1.5}%
\adjustbox{max height=\dimexpr\textheight-1.0cm\relax,max width=\textwidth}{
\begin{tabular}{|p{0.18\linewidth}|p{0.30\linewidth}|p{0.30\linewidth}|p{0.10\linewidth}|p{0.14\linewidth}|p{0.15\linewidth}|p{0.10\linewidth}|p{0.10\linewidth}|p{0.10\linewidth}|p{0.10\linewidth}|}
\hline
\textbf{Process }                                    & $\pom\ \pom\to jjX$                                                   & $\pom\ \reg\!+\!\reg\ \pom\to jjX$ & $\reg\ \reg\to jjX$ & $\pom\ \pom\to jjX$    & $\pom\ \reg\!+\!\reg\ \pom\to jjX$ & $\reg\ \reg\to jjX$\\
\hline
\textbf{Acceptance}                                  & \multicolumn{3}{c|}{$p_{T}(j_{1},j_{2})>20$ GeV }                    & \multicolumn{3}{c|}{$p_{T}(j_{1},j_{2})>50$ GeV }    \\
\hline 
$ \xi_{1,2} < 0.17$  & $3.34\times 10^{4} $ [$2.88\times 10^{4} $, $4.42\times 10^{4} $]      & $1.56\times 10^{4}$ [$1.41\times 10^{4}$, $1.68\times 10^{4} $] & 1610  & 1489 [1198,1829] & 697[594,771] & 72 \\
$0.015 < \xi_{1,2} < 0.17$                           & $2.19\times 10^{4} $  [$1.99 \times 10^{4} $, $2.79\times 10^{4} $] & $1.26\times 10^{4} $ [$1.16\times 10^{4}$, $1.35\times 10^{4}$]                          & 1590  &1030 [876,1269] &576 [536,644] & 70 \\
$0.10 < \xi_{1,2} < 0.17$                            & 2530 [2319,3193]                                                    & 2802 [2627,2850]                                              &  680  & 120[113,135] &148 [140,174] & 35\\
$0.10 < \xi_{1,2} < 0.17\ [\star]$                         &  544 [499,736]                                                       &  865 [813,877]                                                 &  312  &20.5 [10,23] &42 [36,52] & 30 \\
\hline  
\end{tabular}
}
\caption{Number of central diffractive di-jet events for an integrated luminosity of $1$ pb$^{-1}$, and different kinematical windows. For the Pomeron process, the left values inside the brackets stand for $\nu=0.5$, whereas the right values stand for $\nu=-0.5$. The last line $[\star]$ is for $p_{T}$(proton) $>$ 0.4 GeV, instead of the default value 0.15 GeV.}
\label{neventsptcut1}
\end{table}

\begin{figure}[t]
\includegraphics[width=0.49\textwidth]{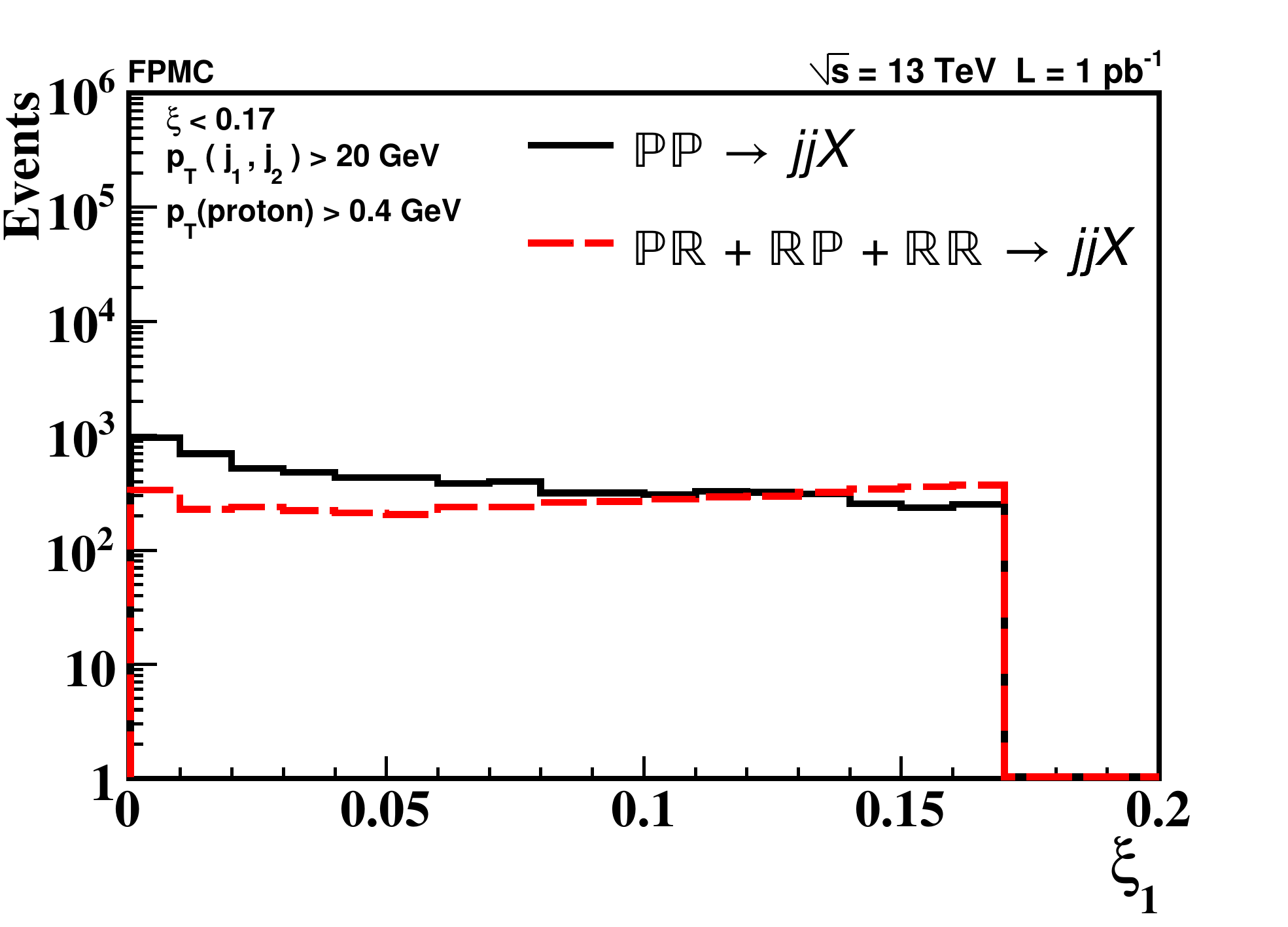}
\includegraphics[width=0.47\textwidth]{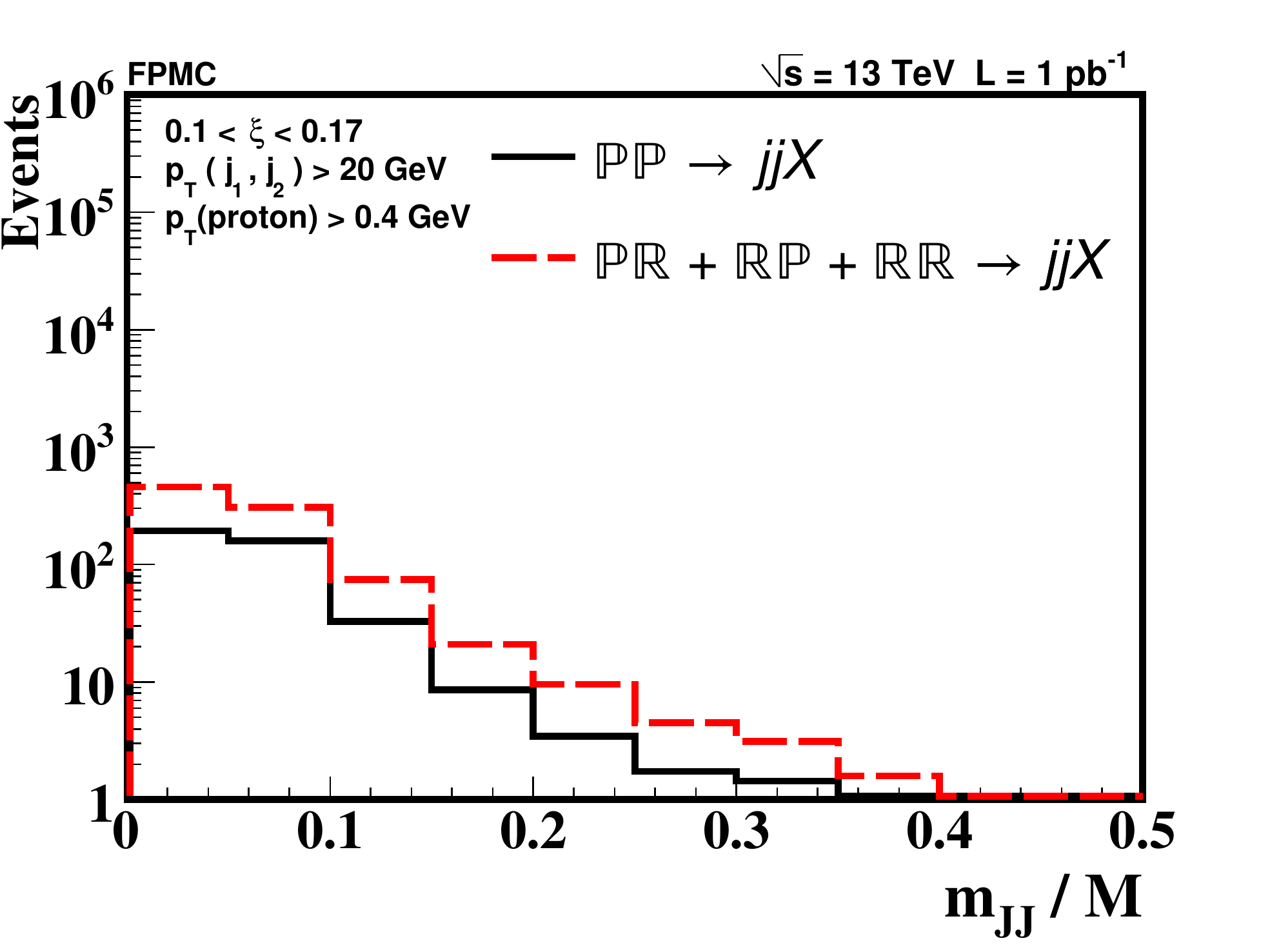}
\caption{
Number of central diffractive di-jet events as function of $\xi_1$ for $\xi_2<0.17$ (left plot) and di-jet mass fraction distribution for $0.1<\xi_{1,2}<0.17$ (right plot), with $p_{T}(j_{1,2}) > 20$ GeV and p$_{T}$(proton) $>$ 0.4 GeV. Increasing the last cut enhances the sensitivity to the Reggeon contribution.}
\label{newprotonptcut}
\end{figure}

Finally, in Figure~\ref{newprotonptcut} we study the sensitivity of our results with respect to the cut on the proton transverse momentum. By choosing an alternative cut of $0.4$ GeV, we are able to increase the sensitivity to the Reggeon contribution, such that near the edge of the proton detector acceptance, it becomes clearly dominant. 
By measuring these distributions, we should be able to study the Reggeon contribution in the LHC data. 

Our results show that di-jets in central diffractive events (formerly known as double-Pomeron-exchange events) at the LHC could be used to study the Reggeon contribution to hard diffractive processes, since a kinematic window of dominance has been identified which could be used experimentally to isolate and constrain it. 

\section{Conclusions}

In this letter, we studied hard diffractive processes in hadron-hadron collisions using the resolved-Pomeron model \eqref{colfact1} and \eqref{colfact2}, supplemented with a Reggeon term according to formula \eqref{dpdfs}. Our goal is to verify whether or not this contribution can be safely neglected at LHC energies. For the moment, it is often ignored when estimating hard diffractive cross-sections in hadron-hadron collisions, even though it is needed for a quantitative description of the diffractive DIS HERA data. The Pomeron structure used in the resolved-Pomeron model is extracted from DIS, therefore a consistency check is in order at the LHC.

To do this, we chose to analyze the diffractive di-jet process at the LHC, assuming an integrated luminosity of $1$~pb$^{-1}$. We have assumed a simple model in which the parton content of the Reggeon $f_{a/\reg}(\beta,\mu^2)$ is given by the pion structure function, but it should be pointed out that the related uncertainties are large since the Reggeon structure at low $\beta$ and high transverse momentum scales is essentially unknown and unconstrained experimentally. 

Our calculations have been performed using the Forward Physics Monte Carlo program. In the case of single diffractive di-jet production, our results confirm the expectation that the Reggeon contribution is comparable to the Pomeron contribution only for $\xi \gtrsim 0.1$. Since the acceptance of the LHC forward proton detectors can go up to $\xi \sim 0.15 - 0.2$, it must be carefully taken into account when the total diffractive mass becomes large, which is the case for a number of final states considered in the literature, e.g.~\cite{N.Cartiglia:2015gve}.

In the case of central diffractive di-jet production, we find that Reggeon exchanges contribute much more, and can almost never be completely ignored, at least in our model.
For large values of $\xi_{1,2}$ but still within the detector acceptances, processes involving Reggeons can even dominate over the double-Pomeron exchange. This should allow relatively clean experimental studies in order to better constrain the Reggeon parton content and correct the model. Subsequently, many phenomenological studies of double-Pomeron-exchange events at the LHC, such as \cite{Marquet:2012ra,Marquet:2013rja,Kohara:2015nda}, will have to be corrected in order to take into account the possibility to exchange Reggeons as well.

\begin{acknowledgments}

D.E. Martins, M. Rangel and A.V. Pereira acknowledge the financial support from the funding agencies CNPq, CAPES and FAPERJ (Brazil). D.E. Martins thanks the Centre de Physique Th\'eorique of \'Ecole Polytechnique for hospitality. He also warmly thanks Gregory Soyez for explanations about the FastJet implementation and the related technical aspects on this analysis. 

\end{acknowledgments}


\begin{thebibliography}{10}
\bibliographystyle{unsrt}

\bibitem{Abachi:1994hb}
  S.~Abachi {\it et al.} [D0 Collaboration],
  Phys.\ Rev.\ Lett.\  {\bf 72} (1994) 2332.
  
\bibitem{Abe:1994de}
  F.~Abe {\it et al.} [CDF Collaboration],
  Phys.\ Rev.\ Lett.\  {\bf 74} (1995) 855.

\bibitem{Ingelman:1984ns}
G.~Ingelman and P.~E.~Schlein,
Phys.\ Lett.\ B {\bf 152} (1985) 256.

\bibitem{Collins:1997sr} 
  J.~C.~Collins,
  Phys.\ Rev.\ D {\bf 57} (1998) 3051 
  [Erratum-ibid.\ D {\bf 61} (2000) 019902].

\bibitem{Affolder:2000vb}
  T.~Affolder {\it et al.} [CDF Collaboration],
  Phys.\ Rev.\ Lett.\  {\bf 84} (2000) 5043.


\bibitem{sgap1}
    V.~A.~Khoze, A.~D.~Martin and M.~G.~Ryskin,
    Int.\ J.\ Mod.\ Phys.\ A {\bf 30} (2015) no.08, 1542004. 

\bibitem{sgap2} 
    E.~Gotsman, E.~Levin and U.~Maor,
    Int.\ J.\ Mod.\ Phys.\ A {\bf 30} (2015) no.08, 1542005. 

\bibitem{sgap3}
    B.~Kopeliovich, R.~Pasechnik and I.~Potashnikova,
    Int.\ J.\ Mod.\ Phys.\ E {\bf 25} (2016) no.07, 1642001. 

\bibitem{sgap4} 
    C.~O.~Rasmussen and T.~Sjöstrand,
    JHEP {\bf 1602} (2016) 142. 

\bibitem{Luszczak:2014mta} 
  M.~Luszczak, A.~Szczurek and C.~Royon,
  JHEP {\bf 1502} (2015) 098. 
\bibitem{Luszczak:2014cxa} 
   M.~Łuszczak, R.~Maciuła and A.~Szczurek,
   Phys.\ Rev.\ D {\bf 91} (2015) no.05, 054024. 
\bibitem{Luszczak:2016csq} 
   M.~Luszczak, R.~Maciula, A.~Szczurek and M.~Trzebinski,
   arXiv:1606.06528 [hep-ph].


\bibitem{N.Cartiglia:2015gve}
  e.~N.Cartiglia {\it et al.} [LHC Forward Physics Working Group Collaboration],
  CERN-PH-LPCC-2015-001, SLAC-PUB-16364, DESY-15-167.

\bibitem{fpmc} 
  M.~Boonekamp, A.~Dechambre, V.~Juranek, O.~Kepka, M.~Rangel, C.~Royon and R.~Staszewski,
  arXiv:1102.2531 [hep-ph].

\bibitem{dglap}
  G.~Altarelli and G.~Parisi,
  Nucl.\ Phys.\  B {\bf 126} (1977) 298;\\
  V.~N.~Gribov and L.~N.~Lipatov,
  Sov.\ J.\ Nucl.\ Phys.\  {\bf 15} (1972) 438;
  Sov.\ J.\ Nucl.\ Phys.\  {\bf 15} (1972) 675;\\
  Y.~L.~Dokshitzer,
  Sov.\ Phys.\ JETP {\bf 46} (1977) 641.
     
\bibitem{Aktas:2006hy} 
  A.~Aktas {\it et al.}  [H1 Collaboration],
  Eur.\ Phys.\ J.\ C {\bf 48} (2006) 715.

\bibitem{Aktas:2006hx} 
  A.~Aktas {\it et al.} [H1 Collaboration],
  Eur.\ Phys.\ J.\ C {\bf 48} (2006) 749.
 
\bibitem{herwig}
G.~Corcella, I.~G.~Knowles, G.~Marchesini, S.~Moretti, K.~Odagiri, P.~Richardson, M.~H.~Seymour and B.~R.~Webber,
arXiv:0210213  [hep-ph].

\bibitem{Trzebinski:2014vha}
  M.~Trzebi\'nski,
  Proc.\ SPIE Int.\ Soc.\ Opt.\ Eng.\  {\bf 9290} (2014) 929026.

\bibitem{fastjet:2012}  
  M.~Cacciari, G.~P.~Salam and G.~Soyez,
  Eur.\ Phys.\ J.\ C {\bf 72} (2012) 1896.
 
 \bibitem{Kepka:2007nr}
  O.~Kepka and C.~Royon,
  Phys.\ Rev.\ D {\bf 76} (2007) 034012.
 
\bibitem{Khoze:2000cy}
  V.~A.~Khoze, A.~D.~Martin and M.~G.~Ryskin,
  Eur.\ Phys.\ J.\ C {\bf 14} (2000) 525.

\bibitem{Khoze:2008cx}
  V.~A.~Khoze, A.~D.~Martin and M.~G.~Ryskin,
  Eur.\ Phys.\ J.\ C {\bf 55} (2008) 363.
    
\bibitem{Kaidalov:2001iz}
  A.~B.~Kaidalov, V.~A.~Khoze, A.~D.~Martin and M.~G.~Ryskin,
  Eur.\ Phys.\ J.\ C {\bf 21} (2001) 521.
  
\bibitem{Bartels:2006ea}
  J.~Bartels, S.~Bondarenko, K.~Kutak and L.~Motyka,
  Phys.\ Rev.\ D {\bf 73} (2006) 093004.
  
\bibitem{Luna:2006qp}
  E.~G.~S.~Luna,
  Phys.\ Lett.\ B {\bf 641} (2006) 171.
  
\bibitem{Frankfurt:2006jp}
  L.~Frankfurt, C.~E.~Hyde, M.~Strikman and C.~Weiss,
  Phys.\ Rev.\ D {\bf 75} (2007) 054009.
  
\bibitem{Gotsman:2007ac}
  E.~Gotsman, E.~Levin and U.~Maor,
  arXiv:0708.1506 [hep-ph].
  
\bibitem{Gotsman:2011xc} 
  E.~Gotsman, E.~Levin and U.~Maor,
  Eur.\ Phys.\ J.\ C {\bf 71} (2011) 1685. 
  
\bibitem{Achilli:2007pn}
  A.~Achilli, R.~Hegde, R.~M.~Godbole, A.~Grau, G.~Pancheri and Y.~Srivastava,
  Phys.\ Lett.\ B {\bf 659} (2008) 137.
  
\bibitem{Chatrchyan:2012vc} 
  S.~Chatrchyan {\it et al.} [CMS Collaboration],
  Phys.\ Rev.\ D {\bf 87} (2013) no.01, 012006. 
  
\bibitem{Aad:2015xis} 
  G.~Aad {\it et al.} [ATLAS Collaboration],
  Phys.\ Lett.\ B {\bf 754} (2016) 214. 
   
\bibitem{Aad:2012pw}
  G.~Aad {\it et al.} [ATLAS Collaboration],
  Eur.\ Phys.\ J.\ C {\bf 72} (2012) 1926.

\bibitem{Marquet:2012ra}
  C.~Marquet, C.~Royon, M.~Trzebi\'nski and R.~\v{Z}leb\v{c}\'ik,
  Phys.\ Rev.\ D {\bf 87} (2013) no.03, 034010.

\bibitem{Marquet:2013rja}
  C.~Marquet, C.~Royon, M.~Saimpert and D.~Werder,
  Phys.\ Rev.\ D {\bf 88} (2013) no.07,  074029.

\bibitem{Kohara:2015nda}
  A.~K.~Kohara and C.~Marquet,
  Phys.\ Lett.\ B {\bf 757} (2016) 393.

\end{thebibliography}
\end{document}